\documentclass[apjl]{emulateapj}

\usepackage{mathptmx}

\begin{document}

\title{Heating by Acoustic Waves of Multiphase Media}
\author{Doron Chelouche\altaffilmark{1,2}}
\altaffiltext{1} {School of Natural Sciences, Institute for Advanced Study,
          Einstein Drive, Princeton 08540, USA; doron@ias.edu}
\altaffiltext{2} {Chandra Fellow}
\shortauthors{Chelouche D.}
\shorttitle{Acoustic Heating}

\begin{abstract}

We study the emission and dissipation of acoustic waves from cool dense clouds in
pressure equilibrium with a hot, volume-filling dilute gas component. In our
model, the clouds are exposed to a source of ionizing radiation whose flux level varies with 
time, forcing the clouds to pulsate. We estimate the rate at which acoustic 
energy is radiated away by an ensemble of clouds and the rate at which it is absorbed by,
and dissipated in, the hot dilute phase. We show that acoustic energy can be a 
substantial heating source of the hot gas phase when the mass in the cool component
is a substantial fraction of the total gas mass. We investigate the applicability of our
results to the multiphase media of several astrophysical systems, including quasar outflows and cooling 
flows. We find that acoustic heating can have a substantial effect on the thermal properties of the hot phase in those systems.

\end{abstract}


\section{Introduction}

Multi-thermal-phase gas configurations are often encountered in astrophysics. For
example, interstellar gas is known to consist of several
thermally distinct components whose temperatures are in the range $10^2-10^6$\,K,
(e.g, McKee \& Ostriker 1977). Gaseous halos of $L^*$ galaxies constitute 
cool ($10^4$\,K) gas condensations embedded in a hot ($10^6$\,K) volume-filling virialized 
gas (e.g., Mo \& Miralda-Escude 1996, Chelouche et al. 2007 and references therein). 
The intracluster medium (ICM) of nearby clusters shows rich structure of cool ionized filaments
embedded in a hot X-ray emitting material and extending to large scales (e.g., Conselice et al. 2001, Fabian et al. 2003). A similar structure
is also observed around radio-loud quasars (RLQs) where a considerable fraction of the mass of the gaseous
nebulae  lies in cool clumps of gas (e.g., Crawford \& Fabian 1992, Fu \& Stockton 2006). More recently, it has been realized that many astrophysical outflows are also multiphase with cool condensed material embedded in a hotter  and more dilute ambient medium. These systems include: planetary nebulae (e.g., Meaburn et al. 1992), stellar winds (e.g., Bouret et al. 2005), and quasar flows (e.g., Das et al. 2005).

The physical mechanisms responsible for multiphase gas configurations in different 
systems are quite diverse. A partial list includes thermal instability and 
compression by radiative shocks in hot material,  and evaporation and gravitational binding
of cool material. While our understanding of the complex physics leading to multiphase 
structures is incomplete, it is well established that such systems are a 
wide-spread phenomenon.

Recent studies suggest that some multiphase systems seem to suffer from a
heat "deficit". One example is that of highly ionized X-ray outflows
from the central regions of Seyfert galaxies that are believed to be
thermally driven and are seen to reach high velocities despite the
effect of adiabatic cooling (e.g., Chelouche \& Netzer 2005). An 
analogous case to that are the large scale narrow line region (NLR) outflows in
active galactic nuclei (AGN) where the effect of adiabatic cooling is even more
pronounced (e.g., Everett \& Murray 2007).  These discrepancies may be 
alleviated if a yet unidentified heating source is lurking in those systems which would balance adiabatic losses. 
Another example is that of cooling flows in which the time for 
radiative cooling of the hot component
appears to be short compared to the Hubble time, yet the
gas does not  cool significantly. Various explanations have been put forward to
explain the cooling flow problem including thermal conduction (e.g., Zakamska \&
Narayan 2003), $PdV$ work done by expanding radio bubbles that are inflated by 
(recently activated) AGN (e.g., Fabian et al. 2003 and references
therein; see also Mathews et al. 2006), and the dissipation of gravitational energy via cool condensations falling through 
the hot gas (Murray \& Lin 2004; see also Dekel \& Birnboim 2007).

In this paper we wish to address the general problem of the heating/cooling balance in
multiphase media around sources with time-varying ionizing fluxes.  Flux variability
is a characteristic of many astronomical objects and of active (accreting) systems in particular [see e.g., Geha et al. (2003) for the case of quasars and Zezas et al. (2007) for a study of X-ray binaries].  
While the examples given in this paper focus on  quasar-related phenomena, we emphasize that the general processes outlined here may apply to other systems such as  X-ray binaries and variable stars.

This paper is organized as follows: In  \S2 we present a general analytic formalism that describes the radiative forcing of clouds which causes them to pulsate and emit acoustic waves into their surrounding medium. Absorption of acoustic radiation and several heat transport mechanisms are also discussed. We address a few specific systems to which this physical mechanism may apply in \S3. Numerical calculations confirming some aspects of our analytic approximations are also presented. Our conclusions follow in \S4.

\section{The Emission and Dissipation of Acoustic Waves}

\subsection{General Setup}

Here we consider a generic configuration of a multiphase medium by
assuming dense and cool gas condensations (clouds) with time-averaged
temperature $T_c$ embedded in a hot, dilute, volume-filling medium
at temperature $T_h$ (see Fig. 1). We assume that the two phases are, in a
time-average sense, in pressure equilibrium such that
\begin{equation}
\rho_h=\rho_c \frac{T_c}{T_h},
\end{equation}
where $\rho_c,~\rho_h$ are the densities of the cool and hot
component, respectively.  Characterizing such an equilibrium
requires knowledge of the relevant heating and cooling mechanisms in
each phase. These processes are complicated, but we note that, generally, cooling is $\propto
\rho^2$; hence the cooling time would be shorter for the dense and cool phase
compared to the dilute and hot one. In a steady-state thermal balance, this is also true for the heating rates.  

Below we show that a fraction of the photon energy which is absorbed by the cool gas can be redirected to heat the hot ambient medium by means of  acoustic waves.  Such
a process naturally occurs when the cool phase is photo-heated by a
time-varying radiation field (e.g.,  that of a quasar). Clouds
distributed throughout the volume of the system  would change their
pressure due to the varying radiative heating and would
therefore pulsate. As such, they would act as bells (or monopole loudspeakers) emitting
acoustic radiation into their environment. Unlike Murray \& Lin (2004), our model for the heating of the volume-filling hot  medium does not require differential velocities 
between the phases to operate, and does not necessarily lead to the grinding of the cool phase.

\begin{figure}
%includegraphics[width=7in]{fig0.eps} 
\plotone{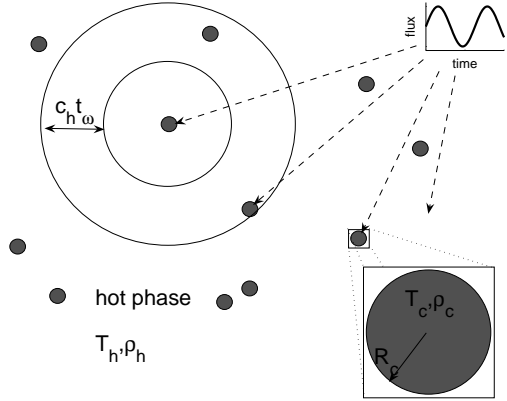} 
\figcaption{A schematic (not-to-scale) view of the multiphase medium considered in this work. Clouds (dark circles) are embedded in a hot dilute medium and are roughly in pressure equilibrium with it (see text). The clouds are assumed to be spherical with a radius $R_c$, density, $\rho_c$, and temperature, $T_c$ (see inset). The entire volume is illuminated by an ionizing source of  varying flux (upper-right corner) that radiatively forces the clouds  to pulsate and emit acoustic waves into their surrounding medium.} 
\label{geometry}
\end{figure}

\subsection{Emission}

Consider an isolated spherical cloud which is exposed to a source of
ionizing flux that varies with time $t$ (Fig. 1). Consider also a specific mode of
oscillation such that the flux, 
\begin{equation}
F(t)=F_0+\delta F e^{i\omega t},
\end{equation}
where $t_\omega = 2\pi \omega^{-1}$ and $\delta F$ is the amplitude of variation. We further assume that
\begin{equation}
{\rm max} ( t^c_{\rm cool}, t^c_{\rm heat}) < t_\omega \ll  {\rm max} ( t^h_{\rm cool}, t^h_{\rm heat})
\end{equation}
with $t^c_{\rm cool},~t^c_{\rm heat}~(t^h_{\rm cool},~t^h_{\rm heat})$ being the cooling time and heating time of the cool (hot) phase, respectively (see \S 3 for the discussion of specific systems). In this regime, the cloud reacts instantaneously to the varying flux by changing its temperature, and hence its pressure (we check the validity of this condition for specific systems in \S 3). So, for example, as the level of ionizing flux rises, more energetic electrons are injected into the gas thereby  increasing its temperature.  The cloud, being over-pressured compared to its environment, would
then expand into the ambient medium. For large enough flux variations ($\delta F/F\sim 1$),  the cloud's temperature and pressure ($P$) variations satisfy $\Delta T_c/T_c \sim \Delta P/P\sim 1$ (see \S 3 for detailed photoionization calculations). In this case, the cloud will expand at roughly its sound speed, $c_c$. We neglect any time dependence of $c_c$ in the present analysis as this is proportional to $\sqrt{\Delta P/P}$ which is of order unity. We note that since $t_\omega \ll t^h_{\rm cool}$, the hot phase will not react to flux variations of this duration and its properties may be considered constant over $t_\omega$. The last condition may be relaxed if the hot component is in a range of parameter space where $\partial {\rm ln}(T)/\partial {\rm ln} (F)=0$, as would be the case for gas at the Compton temperature. 

In what follows, we shall work in the long wavelength regime, namely,
$t_\omega \geq R_c/c_h$ where $R_c$ is the cloud radius (see below for a more general definition of $t_\omega$ which takes into account the finite cooling time of the cloud). At much higher frequencies a phase develops between the gas pressure
and velocity (as they become orthogonal on small scales) so that the cycle-averaged radiation is small.

Acoustic flux is often over-estimated in the literature. When a cloud
$-$ or a piston $-$ expands subsonically (indeed $c_c\ll c_h$ by
construction in our case), the ambient medium in its immediate
environment would be effectively incompressible and would 
pulsate {\it with} the cloud not producing sound (i.e., compression)
waves. Sound waves {\it are} produced once the surrounding medium has
not had time to respond to the changing volume of the region interior
to it which occurs at a distance $\sim c_ht_\omega$ from the
cloud. This distance marks the transition to the radiative zone (see
appendix). From the continuity condition, the velocity at that location
\begin{equation}
\dot{\xi} = c_c \left (1+ \frac{ c_h t_\omega}{R_c}\right )^{-2}.
\end{equation}
Hence, for large enough $t_\omega$, $\dot{\xi}\ll c_c$. The
acoustic luminosity radiated away by a single pulsating cloud would therefore be
\begin{equation}
L_{\rm acoustic}=2\pi R_c^2 \rho_h c_c^2 c_h \left ( 1+ \frac{ c_h t_\omega}{R_c}\right )^{-2}.
\end{equation}
Clearly, for large enough $t_\omega$, $L_{\rm acoustic}$ becomes very small.  

It is interesting to compare the total energy emitted over one cycle
to the energy stored in the hot gas within a shell defined by the
surface of the cloud at a distance $r=R_c$ and $r=R_c+c_ht_\omega$. As
$\dot{\xi}$ declines rapidly with $r$, we can write 
\begin{equation}
\frac{L_{\rm acoustic} t_\omega}{4/3\pi R_c^3 \rho_h c_c^2}\propto \frac{R_c}{c_h t_\omega}.
\end{equation}
Hence, for long enough wavelengths, this ratio is $\ll 1$ and most
of the $PdV$ work done by the cloud by expanding into the ambient
medium is returned to it during the compression phase (e.g., Leighton 1997 and the appendix). This is also why typical loudspeakers are rather inefficient in producing sound converting, on average, only $\sim$1\% of the electric power to audible acoustic power.

The ratio of acoustic energy emitted by all the clouds over the
cooling timescale of the hot phase (from adiabatic or
radiative losses) and the thermal energy stored in the hot phase, $E_h$,
is
\begin{eqnarray}
\displaystyle \frac{E_{\rm acoustic}}{E_h} &  = & \frac{3}{2} \frac{\epsilon_V}{1-\epsilon_V} \left ( \frac{c_c}{c_h} \right )^2 \frac{c_h t^h_{\rm cool}}{R_c} \left ( 1+ \frac{c_h t_\omega}{R_c} \right )^{-2} \\
\nonumber \displaystyle & \sim & \epsilon_V \left ( \frac{c_c}{c_h} \right )^2 \frac{R_c}{c_h t_\omega} \frac{t^h_{\rm cool}}{t_\omega} 
\end{eqnarray}
with $\epsilon_V$ being the volume filling factor of the cool
phase. The last expression is valid in the limit  $c_h t_\omega /R_c \gg 1$ and $\epsilon_v \ll 1$. Clearly, a pre-requisite for effective acoustic heating is that $E_{\rm
acoustic} \geq E_h$. This is achieved for short $t_\omega$. Nevertheless, $t_\omega$ cannot be arbitrarily short since it must satisfy
\begin{equation}
t_\omega \geq t_\omega^0={\rm max} \left ( \frac{R_c}{c_h}, \frac{\rho_c c_c^2}{ \Lambda (T,\rho_c,F)} \right )
\end{equation}
where $\Lambda$ is the net cooling/heating rate per unit volume averaged over half a cycle. If the converse is true then the clouds'  thermal state would change little with time and radiative forcing would be less efficient (this is shown to be the case in \S3). For $t_\omega > t_\omega^0$, $E_{\rm acoustic}/E_h \propto t_\omega^{-2}$, hence $t_\omega^0$ is the mode which is most efficient in converting photon flux to acoustic flux.

It is interesting to compare the acoustic luminosity emitted by the cloud to its photo-heating rate
\begin{equation}
\frac{L_{\rm acoustic}}{4\pi/3 R_c^3 \Lambda} \simeq \frac{R_c}{c_ht_\omega } \left (  \frac{c_c}{c_h} \right )^2 \frac{ \rho_c c_c^2}{\Lambda t_\omega} \leq \left ( \frac{ c_c}{c_h} \right )^2.
\end{equation}
Clearly, only a small fraction of the photo-heating luminosity is transformed to acoustic luminosity (e.g., at most $\sim$1\% will be converted to acoustic luminosity for $T_c=10^4$\,K and $T_h=10^6$\,K).

For the case in which the emitted acoustic radiation is fully absorbed in the system  then acoustic heating would overcome the cooling  of the hot phase once 
\begin{equation}
\frac{\epsilon_V}{1-\epsilon_V} \frac{\Lambda(\rho_c)}{\Lambda (
\rho_h)} \left ( \frac{c_c}{c_h} \right )^2 \sim 1,
\end{equation} 
where we have used equation 9 and our assumption of pressure equilibrium.
If the ratio of the cooling rates for the hot and cold phases is $\sim (\rho_h/\rho_c)^2$ (as would be the case if, for example, the thermal properties of the hot phase are determined primarily by photo-heating; see \S 3)  then we require that the mass fraction of the cool component, $\epsilon_M$, satisfies 
\begin{equation}
\epsilon_M \sim \frac{\rho_c}{\rho_h}\epsilon_V \sim 1.
\end{equation}
This brings us to an important conclusion:  {\it the mass of the cool phase should be at least comparable to that of the hot phase for acoustic heating to be effective}.  We discuss this result more quantitatively in \S 3.1-3.3.

It is possible to analytically estimate the typical size of clouds that can give rise to significant acoustic heating. A natural scale comes from the requirement for a minimal $t_\omega^0$ which occurs for $R_c/c_h=\rho_c c_c^2/\Lambda$ (see equation 8 and the numerical results of \S3). This gives the following order-of-magnitude estimate
\begin{equation}
{R_c} \simeq \frac{t^c_{\rm cool}}{l/c_h}l,
\end{equation}
i.e.,  $R_c/l$ is of order the ratio of the cooling time of the cool phase to the sound-crossing timescale of the hot phase. A specific case worth mentioning is that of adiabatically cooling/heating, thermally-driven systems (such as X-ray and NLR quasar outflows). Here, the thermal state of the hot medium satisfies
\begin{equation}
P\sim \Lambda \frac{l}{c_h}.
\end{equation}
That is, the time it take for the entire volume to heat (or cool) is of order the dynamical time. Using this property of such systems, we can readily estimate estimate $R_c$  to be,
\begin{equation}
R_c \sim \left ( \frac{c_c}{c_h} \right )^4l = \left ( \frac{\rho_h}{\rho_c} \right )^2l
\end{equation}
[in this case, the required number of clouds in the system is roughly $(\rho_c/\rho_h)^5$]. We emphasize, however, that this provides only a rough estimate for $R_c$ and that, in fact, there is a range of cloud sizes which can induce significant acoustic heating, as we demonstrate for a few specific systems in \S3.

\subsection{Absorption}

The  emission of acoustic
energy by a cool cloud does not guarantee that the energy will dissipate over the relevant spatial scales required to effectively heat up the
volume. Dissipation of acoustic energy in the hot medium occurs via absorption (and
conductance) as well as by steepening of acoustic waves into weak
shocks. The latter process is negligible in our case since the clouds pulsate 
subsonically and the wave amplitude is quickly diminished in the spherical case (c.f., Stein \& Schwartz 1972).

The relevant dissipation length scale due to absorption of radiation is (e.g., Mathews et al. 2006)
\begin{equation}
l_d\simeq \mu^{-1}P c_h \left ( \frac{t_\omega}{2\pi} \right) ^2, 
\label{ldiss_a}
\end{equation}
where $\mu$ is the {\it effective} viscosity (including both the
effects of viscosity and conduction; e.g., Fabian et al. 2005). Here
we take the usual Braginskii-Spitzer (e.g., Spitzer 1962) value for
high temperature solar composition gas $\sim 10^{-16} \eta
T_h^{5/2}~{\rm g~cm^{-1} s^{-1}}$ (e.g., Lang 1999) which is
applicable to e.g., the ICM for $\eta=10^{-2}-1$ 
(Narayan \& Medvedev 2001,
Fabian et al. 2005; but see Ettori \& Fabian 2000 for possibly much
lower values). Due to the lack of  observational constraints on the 
conductivity of other systems  (e.g., quasar outflows), we shall
assume values similar to those of the  ICM.

Rewriting equation 7 with dissipation included, we obtain a generalized condition for the importance of acoustic heating,
\begin{equation}
\frac{E_{\rm acoustic}^{\rm dissipated}}{E_h}=\frac{E_{\rm acoustic}}{E_h} \left ( 1-e^{-l/l_d} \right ) \gtrsim 1.
\end{equation} 
For $l_d \gg l$, absorption is inefficient and most of the acoustic energy (a fraction $\sim 1-(l/l_d)$ of it) will escape the system. Absorption is  efficient over the entire volume  for $l_d \sim l$. For $l_d \ll l$, absorption is very efficient yet is localized to the environment of the clouds. In this case several additional conditions are required to effectively heat up the entire volume. These are discussed below.

\subsubsection{The case of $l_d \ll l$}

For volumetric heating to be important, we  require many clouds which are homogeneously distributed throughout the volume of the system and/or the presence of some heat transport mechanism. For clouds distributed randomly in space (as indeed seems to be observationally supported; see \S 1), the characteristic distance between clouds is $\simeq \epsilon_V^{-1/3}R_c$. If no heat transport mechanism is operating then we require 
\begin{equation}
\epsilon_V^{1/3}\frac{l_d}{R_c} \geq 1
\end{equation} 
so that dissipation of acoustic waves due to all clouds occurs throughout the volume. 
A much less restrictive case is that in which heat conduction and/or advection are present. In 
this case, even if the above condition is not satisfied, heat could still be distributed over the entire
 volume.

Heat conduction is due to free
electrons whose velocity is, $v_e\simeq \sqrt{3k_BT_e/m_e}\simeq
6\times10^3 \sqrt{T_h/10^6\,{\rm K}}~{\rm km~s^{-1}}$. The propagation of
electrons through a plasma takes the form of a random walk with some
mean free path, $\lambda_e$.  The largest uncertainty arises in estimating $\lambda_e$, which depends on the strength and
configuration of the magnetic field, $B$, as well as on the pressure
and density of the (hot) medium. If $B=0$ then $\lambda_e =
\lambda_e^{B=0} \simeq 10^4T_h^2/(\rho_h/m_H)$\,cm (where $m_H$ is the mass of a hydrogen atom; see Cowie \& McKee
1977). When $B>0$ the medium becomes anisotropic with $\lambda_e$
being of order the Larmor radius across magnetic field lines. For a
detailed discussion of these issues see Lazarian (2006, and references
therein). Here we assume that $10^{-2}<\lambda_e/\lambda_e^{B=0}<1$
(Narayan \& Medvedev 2001). Hot electrons must traverse a
distance of order the typical cloud separation over $t_{\rm cool}^h$,
so that heat could be uniformly distributed throughout the volume. This results in the following condition for efficient conduction:
\begin{equation}
\frac{\epsilon_V^{-2/3} R_c^2}{v_et_{\rm cool}^h \lambda_e} \lesssim 1.
\end{equation}

In addition to conductance, heat may be advected by e.g., turbulent
motion in the gas (we do not consider here the additional heating by
the dissipation of turbulent energy; e.g., Chelouche \& Netzer
2005). For a Kolmogorov-type transonic turbulence, the velocity of eddies of size $l_e$ is
$v_e=c_h(l_e/l)^{1/3}$ and for efficient advection we therefore
require 
\begin{equation}
\frac{\epsilon_V^{-1/3} l^{1/3} R_c^{2/3}}{c_h t_{\rm cool}^h} \lesssim 1.
\end{equation}

\subsection{Caveats}

We have outlined a general analytic framework for studying the
emission of acoustic waves by pulsating cool, dense clouds and
the dissipation of such waves in a hot surrounding medium. We have
not addressed  the issue of cloud longevity, which could be  important
as clouds are prone to various hydrodynamic instabilities such as 
Kelvin-Helmholtz and Rayleigh-Taylor, as well as evaporation/condensation.  
Our understanding of these processes is rather limited in 
the general astrophysical context as well as in the specific systems considered below,
and we make no attempt to further address this here.
That said, the ubiquity of multiphase structures in astrophysical systems implies that 
such configurations are, at least statistically, quasi-steady-state phenomena. 

A further caveat concerns the role of magnetic fields in determining the propagation
of sound waves in the medium. Better estimates for the magnitude of (magneto-)acoustic 
heating will require not only detailed numerical simulations but also a better understanding 
of the individual systems to which this process may apply.

\begin{figure}
\plotone{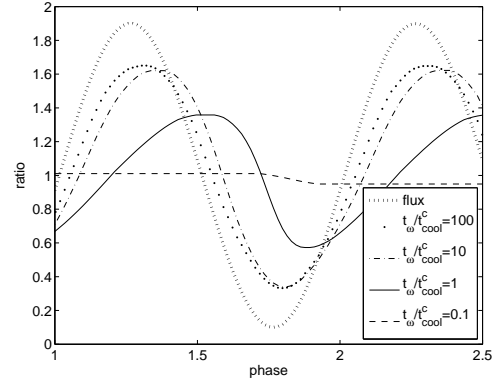} 
\figcaption{Time-dependent calculations of  the temperature of a cloud which is exposed to a time varying flux of a sinusoidal form (denoted by a dotted line; see legend). All quantities are shown relative to their mean values. $t_\omega$ was varied in the range $(0.1-100)\times t^c_{\rm cool}$ as calculated from the steady-state solution. As expected, the cloud thermal state follows the flux variations for $t_\omega \gg t^c_{\rm cool}$. A phase develops and the temperature variation amplitude is somewhat lower for $t_\omega \sim t^c_{\rm cool}$. The cloud's temperature variation is quenched for $t_\omega \ll t^c_{\rm cool}$. These calculations justify our approximation of the heating/cooling rates by their steady-state values (see \S 3).} 
\label{time_d}
\end{figure}

\section{Applications to Astrophysical Systems}

\begin{figure*}
\includegraphics[width=7in]{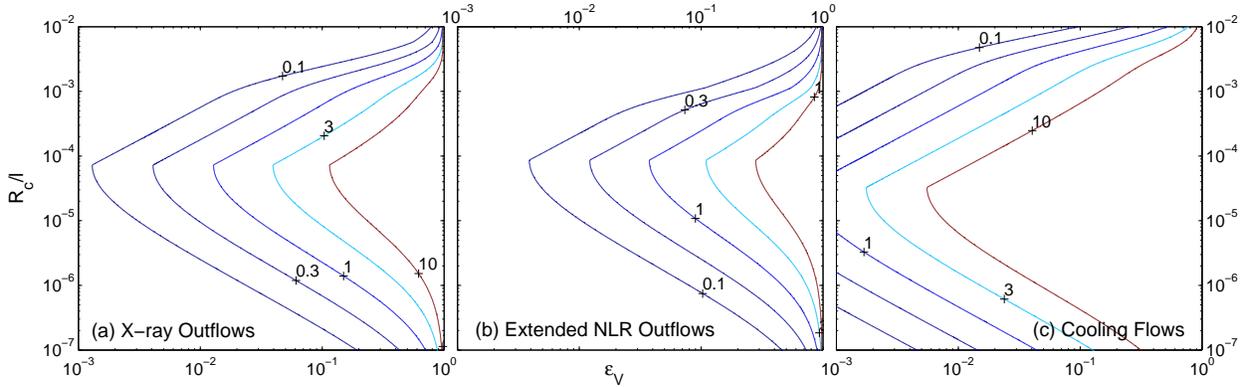} 
\figcaption{Contour level plots of the ratio of acoustic heating to cooling as a function of $\epsilon_V$ and $R_c$ (normalized to the size of the system, $l$) for the systems considered in this paper (\S3.2-3.4). For any $R_c$, it is assumed that there exists a mode with $t_\omega = t_\omega^0$ for which acoustic heating is evaluated. Contours with values $\geq 1$ indicate regions in the parameter space where acoustic heating dominates over cooling in the hot phase. Three astrophysical systems are considered: X-ray outflows in Seyfert I galaxies ({\it left}), extended NLR outflows in Seyfert galaxies ({\it middle}), and cooling flows ({\it right}). Color coding of contours is the same for all panels. We note that the contours' shape
near $\epsilon_V=1$ is somewhat uncertain due to the relatively little volume occupied by the hot material and the importance of absorption and scattering of acoustic waves from the cool phase in this case. Clearly, for all systems there exist a wide range in the parameter space where acoustic heating is important (or even dominant; see text). This  occurs when the cool gas mass is comparable to or larger than that of the hot component. Smaller clouds are more effective in heating the system as long as $R_c/c_h$ is greater than their cooling time. For quasar outflows, the results for the cloud size  and the minimal $\epsilon_V$, where acoustic heating becomes important, are consistent with analytic scalings (see \S 3) and some observational constraints (see text).} 
\label{phase_diagram}
\end{figure*}

In this section we examine how acoustic heating
can be important in  astrophysical environments. It is not
the purpose of this paper to cover all possible cases but rather to study
the applicability of the process to a few specific systems. In all  cases considered 
below there is either an active ionizing source with a varying flux or one 
has recently been active. Here we consider examples  pertaining to quasar 
outflows and to cooling flows. To this end, we have
taken fiducial parameter values from the literature concerning the physical properties
of the cool and hot phase  but note the
appreciable uncertainties associated with them. These parameter values
include rough estimates for $T_c,~T_h$, and $\rho_c$. Estimates for
the total size of the system, $l$, and its cooling time, $t_{\rm
cool}^h$ are also taken from the literature. The characteristic values
of $\epsilon_V$ and $R_c$ are very poorly constrained in those systems  and are
therefore treated as free parameters. 

A major difficulty when estimating the response of  the cloud to radiative forcing is to properly assess the net heating/cooling 
function $\Lambda$ (equation 8) which depends on whether the gas is
in photoionization and thermal equilibrium at all times (as would be the case if
the heating and cooling timescales are the shortest in the problem) or whether its thermal state depends on its history in some intricate way. To keep 
the problem tractable,  we estimate the net heating/cooling function 
by the value of the cooling function under steady-state conditions (in which case its values match those of the heating function). 
For solar metalicity, optically-thin gas exposed to a type-I quasar continuum (Sazonov et al. 2005), the heating rate does not have a strong dependence on the 
temperature, varying by a factor of order unity in the range $10^4<T_c<10^6$\,K, so that an adequate approximation for our purpose is
\begin{equation}
\Lambda (T,\rho)\simeq 10^{-23}(\rho_c/m_H)^2~{\rm erg~s^{-1}~cm^{-3}}.
\end{equation} 
This value shall be used to obtain an order of magnitude estimate for the heating/cooling timescale of the cloud (equation 8).

To check the validity of the above approximation for $\Lambda$ in cases where strict thermal equilibrium is not justified, we have calculated several time-dependent  photoionization models using {\sc cloudy} c07.02 (Ferland et al 1998) and assuming a light-curve, of a fixed amplitude, varying the mode frequency, i.e., $t_\omega$. Ischoric conditions are assumed for the calculations since $c_c/c_h \ll 1$ and we focus on the threshold of the long wavelength regime (see below). Calculations were carried out for the case of $\rho_c/m_H=10^5\,{\rm cm^{-3}}$ and placing the cloud at  1\,pc from the quasar so that its steady-state temperature $-$ if it were exposed to a constant flux $F$ $-$ is $\sim 8\times 10^4$\,K (corresponding roughly to X-ray outflows temperatures; see \S 3.1). The results, shown in Figure. 2, indicate that our estimate for the net cooling function (i.e., $t^c_{\rm cool}$) using steady-state values  is indeed reasonable. In particular, the cloud does not react to the varying flux level of the source  for $t_\omega \ll t^c_{\rm cool}$ and radiative forcing would be inefficient.  The cloud temperature does follow the light-curve for $t_\omega \sim t^c_{\rm cool}$, albeit with a phase, and its temperature fluctuations, $\Delta T_c$, are almost as large as those obtained for $t_\omega \gg t^c_{\rm cool}$ (where little or no phase is present since the cloud reacts instantly to the source's flux level). We note here that $\Delta T_c$ depends not only on $\Delta F$ but also on $\partial {\rm ln}(T)/\partial {\rm ln} (F)$: in particular, for the chosen ionizing continuum, large flux variations are expected for $T_c \sim \,{\rm a~few}\times 10^4-10^5$\,K. The proper treatment of marginally stable or thermally unstable regions (Krolik \& Kriss  2001) is beyond the scope of this paper. 

In the above, we have not addressed the possibility of changes to the spectrum of the ionizing
source which are known to occur (e.g., Netzer et al. 2003) and could influence the gas 
temperature even when the total integrated flux remains unchanged. Such a case can
be easily accommodated within our general formalism (\S 2) by considering $F$ to be  the relevant 
ionizing flux level (e.g., that just above the Lyman edge for $T_c \sim 10^4$\,K gas or around soft X-ray energies for  $T_c \sim 10^5$\,K gas). As spectral variations  of active sources are not well constrained by observations and their inclusion will add little to our understanding of the general effect, we choose to not address this issue here.

Quasars are known to vary considerably ($\delta F/F \sim 1$) over a
wide range of timescales, from minutes to years (e.g., Giveon et
al. 1999, Netzer et al. 2002 and references therein). While current observations do not
directly constrain the power of such variations on the timescales and
wavebands most relevant to our study, it is reasonable to
assume that the ionizing flux of quasars varies considerably over the entire
relevant frequency range.  This motivates us to adopt the following 
approach: we assume that for any choice of system parameters (e.g., 
cloud size, density, and temperature), there exists a mode in the quasar light-curve oscillating with $t_\omega \simeq t_\omega^0$ and having a large amplitude of flux variation ($\delta F/F \sim 1$) . 

Our calculations indicate that, for all systems considered here, uniformly distributed heating 
by acoustic waves is possible so that one (or more) of equations $17-19$ is always satisfied when the more restrictive, equation 16 is.  Generally, smaller clouds (shorter $t_\omega$) occupying a large fraction of the volume give rise to more efficient heating. Nevertheless, as clouds cannot be infinitely small, on account of their cooling time being longer than $t_\omega$ (equation 8), there is a intermediate range of cloud sizes with a total volume above a certain threshold, which is most efficient in heating the ambient medium. The details depend on the system in question as we now discuss.

\subsection{X-ray Outflows in Seyfert 1 Galaxies}

The parameters for X-ray gaseous outflows from Seyfert 1 galaxies which we adopt here are: $T_c=8\times 10^4\,{\rm
K},~T_h=10^7\,{\rm K},~\rho_c/m_H=3\times 10^5~{\rm cm^{-3}},~l=3\,{\rm pc}$
and $t_{\rm cool}^h=10^4$\,years (Chelouche \& Netzer 2005).
Considering equation  16, there is a large range of the
parameter space in which acoustic heating is considerable and
can balance adiabatic cooling (see Fig. 3). In particular, such heating would be effective - that is, comparable to radiative heating -  for $\epsilon_V \gtrsim 10^{-2}$, i.e., once the mass in the cool phase is comparable or larger than that of the hot phase (c.f. equation 11). Cloud sizes should be in the range $10^{14}-10^{15}~{\rm cm}$ ($R_c/l \sim 10^{-4}$; see equation 14) for acoustic heating to be effective. Such values are consistent with full occultation of the X-ray source in Seyfert 1 galaxies (e.g., Chelouche \& Netzer 2005). We note that if $\epsilon_V\simeq 0.1$ (i.e., $\epsilon_M \gtrsim 10$) then acoustic heating may even dominate over photo-heating resulting in a hot-phase temperature that would exceed the Compton temperature. 

\subsection{Extended NLR Outflows in Seyfert Galaxies}

Little is known about the physical properties of NLR
flows, and the existence of a hot medium is mainly inferred from the dynamics of the low temperature,
optically-detected gas. We take fiducial values of $T_c=10^4\,{\rm
K},~T_h=10^6\,{\rm K}$ and consider multiphase gas on $\sim 30$\,pc scales (e.g., Das et al. 2005, Everett \& Murray 2007). For luminosities typical of type-I Seyfert galaxies, the implied density of the hot phase on such scales is of  order $10\,{\rm cm^{-3}}$ (Everett \& Murray 2007; see also Chelouche \& Netzer 2005) hence $\rho_c/m_H=10^3~{\rm cm^{-3}}$. Here, the dynamical time is of order the adiabatic-cooling time which is of order $10^5$\,years (Everett \& Murray 2007). In this case we find a wide range in parameter space where acoustic heating can be important. Everett
\& Murray (2007) quote values for the filling factor of the cool phase as
high as $10^{-2}$. Figure 3 shows that, in this case, the emission of acoustic energy by clouds with $R_c
\sim 10^{16}~{\rm cm}$ could contribute significantly to the heating-cooling balance of the hot phase.  Clouds of this size cannot be individually
resolved by current observations yet similar cloud sizes have been 
inferred from density and column density measurements of UV absorber properties on 
similar scales (e.g., Kraemer et al. 2001 and references therein). Arguments applying to X-ray outflows (equation 18,19) apply here too.

\subsection{Cooling Flows in Galaxy Clusters}

Observations of RLQs show the presence of intrinsically bright 
 O\,II and O\,III emission line regions with apparent sizes of a few$\times
10$\,kpc that consist of numerous unresolved cool clouds (e.g., Fu \& Stockton 2006 and references therein). It
is thought that at least some RLQs and radio galaxies inhabit the
cores of galaxy clusters that possess massive cooling flows
toward their centers (e.g., Crawford \& Fabian 1992). In this case,
the optical emission is thought to originate from cool clouds embedded
within the hot medium (Crawford \& Fabian 1992). It is not known
whether such multiphase gas configurations are a property of all
cooling flows in galaxy clusters; nor is it clear if all cooling flow clusters
which do not cool effectively contain active or recently active nuclei (though they
do seem to have a $>50$\% duty-cycle for jet activity; Fabian \& Sanders 2006). Nevertheless,  we can
use typical cloud parameters inferred from optical
observations and ambient medium parameters deduced from  X-ray studies:
$T_c=2\times 10^4\,{\rm K},~T_h=10^7\,{\rm K},~\rho_c/m_H=10~{\rm
cm^{-3}},~l=5\times 10^4\,{\rm pc}$ and $t_{\rm cool}^h=7\times
10^9$\,years (see e.g., Dunn \& Fabian 2006 and Fu \& Stockton
2006). Figure 3 shows that for $\epsilon_V\gtrsim 10^{-3}$ (indicating 
comparable mass in the hot and cool phase; c.f. Fu
\& Stockton 2006), clouds of size $\sim 0.01-1$\,pc could
efficiently balance radiative cooling in those systems. Interestingly, similar cloud sizes 
 have been deduced by Hamann et al. (2001) for the clouds seen in  absorption toward quasars. 
 Larger values of $\epsilon_V$ (hence $\epsilon_M$) would be required for efficient acoustic 
 heating if the combined  duty-cycle of intracluster active sources is much smaller than unity 
 (Shen et al. 2007). Given the uncertainties associated with the duty-cycle of quasars in cooling flow environments and the properties of the cooling flow itself (e.g., concerning its multiphase structure), it is difficult to assess just how important acoustic heating is, in the form considered here.

\section{Conclusions}

In this paper we have presented a simple model for estimating the effect
of acoustic heating by radiatively-forced pulsating clouds on the thermal state of a hot ambient medium.
We find that smaller clouds
are more efficient acoustic emitters as long as their cooling time
is longer than the oscillation period. Acoustic heating 
is significant when the mass of cool
clouds is comparable to, or larger than that of the hot phase.  In particular, 
for large enough mass in cool gas, acoustic heating may even dominate over
photo-heating. This implies the possible existence of higher temperature gas 
around  sources with variable flux compared to non-variable objects with similar characteristics. 
This work demonstrates that
heating of multiphase photoionized gas may depend not only on the
mean flux of the source but also on deviations from it, as well as on the
structure of the medium itself. We show that there exists a range of clouds sizes which is most efficient in emitting acoustic energy. Applying this model to quasar outflows and
cooling flows, we find that acoustic heating could be important in these
cases if the typical size of cool clouds is of order $10^{-4}$ the size of the system.  Better understanding of the multiphase nature of photoionized environments  in terms
of the relevant densities, cloud sizes, and cooling rates, as well
as the properties of the ionizing source (i.e., its power spectrum of variations) are required to properly 
assess this effect in the broader astrophysical context.

\acknowledgements

This work is supported by NASA through a Chandra Postdoctoral
Fellowship award PF4-50033. It is a pleasure to thank David Bowen, Shane Davis, and Bill Mathews  for a careful reading
and debugging of an earlier version of this paper. We thank
 P. Goldreich, A. K\"{o}nigl, and R. Langer for fruitful discussions, H. Netzer for 
 valuable comments, and G. Ferland and collaborators for the latest version of {\sc cloudy}. Many 
 enjoyable auditioning  hours of both "monopole" and dipole loudspeakers with 
 David Bowen and Matt Kleban are greatly appreciated.

\begin{appendix}

Lagrangian perturbations for the equation of motion of the hot component in spherical coordinates gives,
\begin{equation}
\partial_t^2\xi=c_h^2\nabla(\nabla\cdot\xi)=c_h^2\frac{\partial}{\partial r}\left ( \frac{1}{r^2} \frac{\partial}{\partial r} r^2 \xi \right )
\end{equation}
whose general solution is 
\begin{equation}
\xi=\left [ aj_1(kr)+by_1(kr) \right ]e^{i\omega t},
\end{equation}
where $j_1,~y_1$ are the spherical Bessel functions. The boundary
conditions for the problem are: i) a forced oscillator at the wall of the cloud satisfying 
$\dot{\xi}(r=R_c,t)=c_c {\rm exp}(i\omega
t)$ and ii) a radiatively expanding wave solution at infinity such that $\xi(r\rightarrow \infty)\propto {\rm exp}(i\omega t-kr)$. A solution is obtained upon matching the boundary conditions and is
shown in Figure 4. Clearly, the flux amplitude drops quickly, and by a
large factor, from the surface of the cloud (on the left side) toward
the radiative zone. Hence, most of the energy emitted by the pulsating
cloud at the start of the cycle is returned to it toward the end so
that only a small fraction of the energy is acoustically radiated
away. The acoustic flux at large distances from the cloud surface is
constant in this example since no dissipative mechanism was included in equation
A1.

\begin{figure*}
\centerline{\includegraphics[width=3.0in]{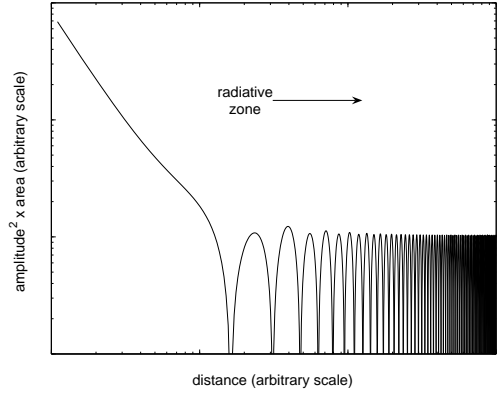} }
%\plotone{fig2.eps} 
\caption{The ``luminosity'', shown here as the amplitude$^2\times$area at some instance $t$, of an outgoing wave driven on the left side with a finite amplitude (a dissipationless wave equation is assumed). Clearly, the amplitude diminishes rapidly toward the radiative zone and the luminosity carried by an outgoing wave is considerably smaller than that which would be {\it naively} calculated at the source of motion (see \S2.2) . Note the logarithmic scales in both axes.} \label{wave}
\end{figure*}

\end{appendix}


\begin{thebibliography}{}

\bibitem[Bouret et al.(2005)]{2005A&A...438..301B} Bouret, J.-C., Lanz, T., 
\& Hillier, D.~J.\ 2005, \aap, 438, 301

\bibitem[Chelouche \& Netzer(2005)]{2005ApJ...625...95C} Chelouche, D., \& 
Netzer, H.\ 2005, \apj, 625, 95

\bibitem[Chelouche et al.(2007)]{2007arXiv0706.4336C} Chelouche, D., 
M{\'e}nard, B., Bowen, D.~V., \& Gnat, O.\ 2007, \apj\ submitted (astro-ph/0706.4336)

\bibitem[Conselice et al.(2001)]{2001AJ....122.2281C} Conselice, C.~J., 
Gallagher, J.~S., III, \& Wyse, R.~F.~G.\ 2001, \aj, 122, 2281

\bibitem[Cowie \& McKee(1977)]{1977ApJ...211..135C} Cowie, L.~L., \& McKee, 
C.~F.\ 1977, \apj, 211, 135

\bibitem[Crawford \& Fabian(1992)]{1992MNRAS.259..265C} Crawford, C.~S., \& 
Fabian, A.~C.\ 1992, \mnras, 259, 265

\bibitem[Das et al.(2005)]{2005AJ....130..945D} Das, V., et al.\ 2005, \aj,  130, 945

\bibitem[Dekel \& Birnboim(2007)]{2007arXiv0707.1214D} Dekel, A., \& 
Birnboim, Y.\ 2007, \mnras\ submitted (astro-ph/0707.1214)

\bibitem[Dunn \& Fabian(2006)]{2006MNRAS.373..959D} Dunn, R.~J.~H., \& 
Fabian, A.~C.\ 2006, \mnras, 373, 959

\bibitem[Everett \& Murray(2007)]{2007ApJ...656...93E} Everett, J.~E., \& 
Murray, N.\ 2007, \apj, 656, 93

\bibitem[Ettori \& Fabian(2000)]{2000MNRAS.317L..57E} Ettori, S., \& 
Fabian, A.~C.\ 2000, \mnras, 317, L57

\bibitem[Fu \& Stockton(2006)]{2006ApJ...650...80F} Fu, H., \& Stockton, 
A.\ 2006, \apj, 650, 80

\bibitem[Fabian et al.(2002)]{2002MNRAS.332L..50F} Fabian, A.~C., Allen, 
S.~W., Crawford, C.~S., Johnstone, R.~M., Morris, R.~G., Sanders, J.~S., \& 
Schmidt, R.~W.\ 2002, \mnras, 332, L50

\bibitem[Fabian et al.(2003)]{2003MNRAS.344L..43F} Fabian, A.~C., Sanders, 
J.~S., Allen, S.~W., Crawford, C.~S., Iwasawa, K., Johnstone, R.~M., 
Schmidt, R.~W., \& Taylor, G.~B.\ 2003, \mnras, 344, L43

\bibitem[Fabian et al.(2005)]{2005MNRAS.363..891F} Fabian, A.~C., Reynolds, 
C.~S., Taylor, G.~B., \& Dunn, R.~J.~H.\ 2005, \mnras, 363, 891

\bibitem[Fabian \& Sanders(2006)]{2006astro.ph.12426F} Fabian, A.~C., \& 
Sanders, J.~S.\ 2006, Proceedings of "Heating vs Cooling in Galaxies and Clusters of Galaxies", August 2006, Garching (astro-ph/0612426)

\bibitem[Ferland et al.(1998)]{1998PASP..110..761F} Ferland, G.~J., 
Korista, K.~T., Verner, D.~A., Ferguson, J.~W., Kingdon, J.~B., \& Verner, 
E.~M.\ 1998, \pasp, 110, 76

\bibitem[Geha et al.(2003)]{2003AJ....125....1G} Geha, M., et al.\ 2003, 
\aj, 125, 1

\bibitem[Giveon et al.(1999)]{1999MNRAS.306..637G} Giveon, U., Maoz, D., 
Kaspi, S., Netzer, H., \& Smith, P.~S.\ 1999, \mnras, 306, 637

\bibitem[Hamann et al.(2001)]{2001ApJ...550..142H} Hamann, F.~W., Barlow, 
T.~A., Chaffee, F.~C., Foltz, C.~B., \& Weymann, R.~J.\ 2001, \apj, 550, 
142

\bibitem[Kraemer et al.(2001)]{2001ApJ...551..671K} Kraemer, S.~B., et al.\ 
2001, \apj, 551, 671

\bibitem[Krolik \& Kriss(2001)]{2001ApJ...561..684K} Krolik, J.~H., \& 
Kriss, G.~A.\ 2001, \apj, 561, 684

\bibitem[Lang(1999)]{1999asfo.book.....L} Lang, K.~R.\ 1999, Astrophysical 
formulae  / K.R.~Lang.~New York : Springer, 1999.~(Astronomy and 
astrophysics library,ISSN0941-7834),

\bibitem[Lazarian(2006)]{2006ApJ...645L..25L} Lazarian, A.\ 2006, \apjl, 
645, L25

\bibitem[Leighton(1997)]{1997asfo.book.....L} Leighton, T.~G.\ 1997, The Acoustic Bubble  / T.G.~Leighton.~Academic Press, 1997.~(ISBN-10: 0124419216),

\bibitem[Mathews et al.(2006)]{2006ApJ...638..659M} Mathews, W.~G., 
Faltenbacher, A., \& Brighenti, F.\ 2006, \apj, 638, 659

\bibitem[McKee \& Ostriker(1977)]{1977ApJ...218..148M} McKee, C.~F., \& 
Ostriker, J.~P.\ 1977, \apj, 218, 148

\bibitem[Meaburn et al.(1992)]{1992MNRAS.255..177M} Meaburn, J., Walsh, 
J.~R., Clegg, R.~E.~S., Walton, N.~A., Taylor, D., \& Berry, D.~S.\ 1992, 
\mnras, 255, 177

\bibitem[Murray \& Lin(2004)]{2004ApJ...615..586P} Murray, S.~D., \& Lin, D.~N.~C.\ 2004, 
\apj, 615, 586

\bibitem[Narayan \& Medvedev(2001)]{2001ApJ...562L.129N} Narayan, R., \& 
Medvedev, M.~V.\ 2001, \apjl, 562, L129

\bibitem[Netzer et al.(2002)]{2002ApJ...571..256N} Netzer, H., Chelouche, 
D., George, I.~M., Turner, T.~J., Crenshaw, D.~M., Kraemer, S.~B., \& 
Nandra, K.\ 2002, \apj, 571, 256

\bibitem[Sazonov et al.(2005)]{2005MNRAS.358..168S} Sazonov, S.~Y., 
Ostriker, J.~P., Ciotti, L., \& Sunyaev, R.~A.\ 2005, \mnras, 358, 168 

\bibitem[Shen et al.(2007)]{2007AJ....133.2222S} Shen, Y., et al.\ 2007, 
\aj, 133, 2222

\bibitem[Spitzer(1962)]{1962pfig.book.....S} Spitzer, L.\ 1962, Physics of 
Fully Ionized Gases, New York: Interscience (2nd edition), 1962,

\bibitem[Stein \& Schwartz(1972)]{1972ApJ...177..807S} Stein, R.~F., \&
Schwartz, R.~A.\ 1972, \apj, 177, 807

\bibitem[Zakamska \& Narayan(2003)]{2003ApJ...582..162Z} Zakamska, N.~L., 
\& Narayan, R.\ 2003, \apj, 582, 162 

\bibitem[Zezas et al.(2007)]{2007ApJ...661..135Z} Zezas, A., Fabbiano, G., 
Baldi, A., Schweizer, F., King, A.~R., Rots, A.~H., \& Ponman, T.~J.\ 2007, 
\apj, 661, 135

\end{thebibliography}
\end{document}